\documentclass[12pt]{iopart}
\newcommand{\be}{\begin{equation}}
\newcommand{\ee}{\end{equation}}
\newcommand{\bea}{\begin{eqnarray}}
\newcommand{\eea}{\end{eqnarray}}
\newcommand{\nn}{\nonumber}
\newcommand{\beas}{\begin{eqnarray*}}
\newcommand{\eeas}{\end{eqnarray*}}
\newcommand{\ben}{\begin{enumerate}}
\newcommand{\enn}{\end{enumerate}}
\newcommand{\slsh}[1]{{\not \! #1}}

\usepackage{graphicx}

\begin{document}

\begin{flushright}
UMSNH-IFM-F-2008-32
\end{flushright}

\title[Gauge Covariance Relations and the Fermion Propagator in MCS-QED$_3$]
{Gauge Covariance Relations and the Fermion Propagator in Maxwell-Chern-Simons QED$_3$}
\author{A Bashir, A Raya and S S\'anchez Madrigal}
\address{Instituto de F\'{\i}sica y Matem\'aticas, Universidad Michoacana 
de San Nicol\'as de Hidalgo, Apartado Postal 2-82, Morelia, Michoac\'an 
58040, M\'exico.}
\ead{raya@ifm.umich.mx, adnan@ifm.umich.mx}

\begin{abstract}
We study the gauge covariance of the fermion propagator in Maxwell-Chern-Simons planar quantum electrodynamics (QED$_3$) 
considering four-component spinors with parity-even and parity-odd mass terms both for fermions and photons. Starting 
with its tree level expression in the Landau gauge, we derive a non perturbative expression for this propagator in an 
arbitrary covariant gauge by means of its Landau-Khalatnikov-Fradkin transformation (LKFT).  We compare our findings 
in the weak coupling regime with the direct one-loop calculation of the two-point Green function and observe perfect 
agreement up to a gauge independent term. We also reproduce results derived in earlier works as special cases of our 
findings. 

\end{abstract}
\pacs{11.15.Tk,11.30.Er,12.20.-m}
\submitto{\JPA}

\maketitle

\section{Introduction}
Gauge symmetry is the cornerstone of our modern understanding of fundamental interactions. At the level of field 
equations,  such a symmetry is reflected in different relations among the Green's functions of a given quantum 
field theory. In quantum electrodynamics (QED), for example, Green's functions verify Ward-Green-Takahashi 
identities~\cite{WGTI}, which relate $(n+1)$-point functions with the $n$-point ones. This set of identities can 
be enlarged by transforming also the gauge fixing parameter $\xi$ to arrive at the Nielsen identities 
(NI)~\cite{Nielsen}. One advantage of these identities over the conventional Ward identities is that 
$\partial /\partial \xi$ becomes part of the new relations involving Green's functions. This fact was 
exploited in \cite{Steele} to prove the gauge independence of some physical observables related to two-point 
Green's functions at the one-loop level and to all orders in perturbation theory. A different set 
of relations, which specify the transformation of Green's functions under a variation of gauge, carry the 
name of Landau-Khalatnikov-Fradkin  transformations (LKFTs) in QED,~\cite{LKF}. LKFTs are non perturbative in nature,  
and hence have the potential of playing an important role in understanding the apparent problems of gauge invariance 
in the strong coupling studies of Schwinger-Dyson equations (SDEs)~\cite{esd}. In this context, the direct 
implementation of LKFTs in SDEs studies has already been reported~\cite{LKFSDE,LKfew}. Gauge dependence studies
of the SDEs must ensure that these transformation for the Green's functions involved are satisfied~\cite{LKfew}
in order to obtain meaningful results. Rules governing LKFTs are better described in coordinate space. 
It is primarily for this reason that some earlier works on its implementation in the study of the 
fermion propagator were carried out in the coordinate  space,~\cite{del1}. Momentum space  calculations 
are more demanding, owing to the complications induced by  Fourier transforms. These difficulties are 
reflected in~\cite{LKFper} and \cite{LKFperb} where the non-perturbative fermion propagator was obtained 
starting from a perturbative one in the Landau gauge in QED in 3 and 4 space-time dimensions.

In this paper, we study QED in three space-time dimensions (QED$_3$) in its general form, taking into account 
parity conserving and violating mass terms both for photons and fermions.  Specific cases of the underlying 
Lagrangian have found many useful applications both in condensed matter physics, particularly in high 
T$_c$-superconductivity and the quantum Hall effect~\cite{Sharapov,supercon,ddw,lg,graphene,zero}, as well 
as in high energy physics, mostly connected to the study of dynamical chiral symmetry breaking and 
confinement, where QED$_3$ provides a popular battleground for lattice and continuum studies~\cite{dcsb}. 
An interesting review of various dynamical effects in (2+1)-dimensional theories with
four-fermion interaction can be found in~\cite{Vshivtsev}.
In all these cases, it becomes a key issue to address the gauge covariance properties of Green's functions. 
We investigate the gauge structure of the 
fermion propagator in the light of the LKFTs. This paper is organized as follows: In the following section, 
considering four-component spinors, we describe the QED$_3$ Lagrangian with parity conserving and violating
mass terms. It leads to a general fermion propagator which we write in a form suitable to study its gauge 
covariance relations. In section, 3 we introduce the LKFT for the fermion propagator and derive the non-perturbative 
expression for the two-point function under consideration. We review some limiting cases of 
our findings, including the massless case, the parity conserving case and the weak coupling expansion, 
which  is  compared against the one-loop calculation of the fermion propagator. It is well known that the parity violation in the
fermion sector radiatively induces a Maxwell-Chern-Simons mass term for the photon. In section 4, we extend our
study to include this case.
At the end, we present our 
conclusions in section 5.

\section{Fermion Propagator}

As compared with its four-dimensional counterpart, only three Dirac matrices are required to describe 
the dynamics of planar fermions. Therefore, one can choose to work with two- or four-component spinors. 
Correspondingly, an irreducible or reducible representation for the $\gamma_\mu$-matrices would
respectively be used.  A discussion on the symmetries of the fermionic Lagrangian with different 
representations of Dirac matrices can be found in~\cite{zero,QED3lag}. In this paper, we work with 
four-component spinors and thus with a $4\times 4$ representation for the Dirac matrices. 
We choose to work in Euclidean space, where the Dirac matrices satisfy the Clifford 
algebra $\{ \gamma_\mu, \gamma_\nu\}=-2\delta_{\mu\nu}$, a realization of which is given by
$$
\gamma_0\,\equiv\,\left(\begin{array}{cc} -i\sigma_3 & 0 \\ 0 & 
i \sigma_3 \end{array}\right)\;,\qquad 
\gamma_1\,\equiv\,\left(\begin{array}{cc} i\sigma_1 & 0 \\ 0 & -i 
\sigma_1 \end{array}\right)\;,\qquad
\gamma_2\,\equiv\,\left(\begin{array}{cc} i\sigma_2 & 0 \\ 0 & -i \sigma_2 
\end{array}\right)\,,
$$
and 
$$
\gamma_3\equiv\left(\begin{array}{cc} 0 & I \\ I & 
0 \end{array}\right)\,,\qquad
\gamma_5\equiv\gamma_0\gamma_1\gamma_2\gamma_3=\left(\begin{array}{cc} 0 & 
-I \\ I & \phantom{-}0 \end{array}\right)\,,
$$
where $\sigma_i,\,i=1,2,3$ are the Pauli matrices and $I$ the $2\times 2$ identity matrix. Notice that 
once we have selected a set of matrices to write down the Dirac equation, say $\{\gamma_0,\gamma_1,\gamma_2\}$, 
two anti-commuting gamma matrices, namely, $\gamma_3$ and $\gamma_5$ remain unused, leading us to
define  two types of chiral-like transformations: $\psi\to e^{i\alpha\gamma_3}\psi\;$ and 
$\psi\to e^{i\alpha\gamma_5}\psi\;.$  Consequently, there exist two types of mass terms for fermions, 
the ordinary $m_e \bar\psi\psi$ and the $m_o \bar\psi\tau\psi\;$ with 
$\tau=\frac{1}{2}[\gamma_3,\gamma_5]=diag(I,-I)$, sometimes referred to as the Haldane mass term.
The former violates chirality, whilst the later  is invariant under chiral transformations. 
Defining parity so that it corresponds to the inversion of only one spatial axis (preserving 
its discrete nature), we can  represent parity transformation by  ${\cal P}=-i\gamma_5\gamma_1$. 
We thus see that $m_e \bar\psi\psi$ is parity invariant but $ m_o \bar\psi\tau\psi$ is not. 
This would justify the use of subscripts $e$ and $o$ for parity-even and parity-odd quantities 
throughout the paper. We shall be working with the Lagrangian:
\be
{\cal L}= \bar\psi (i\slsh{\partial}+e\slsh{A}-m_e-\tau m_o)\psi
-\frac{1}{4}F_{\mu\nu}F_{\mu\nu}-
\frac{1}{2\xi}(\partial_\mu A_\mu)^2, \label{lag}
\ee
where the quantities carry their usual meaning. There are many planar condensed matter models in 
which the low energy sector can be written as this effective form of QED$_3$, for which the 
physical origin of the masses depends on the underlying system~\cite{Sharapov}~: 
$d$-wave cuprate superconductors~\cite{supercon}, $d$-density-wave states~\cite{ddw}, 
layered graphite~\cite{lg}, including graphene in the massless version~\cite{graphene} 
and a special form of the integer quantum Hall Effect without Landau levels~\cite{zero}. 
Chiral symmetry breaking and confinement for particular forms of this Lagrangian~\cite{dcsb} 
and dynamical effects of four-fermion interactions in similar models~\cite{Vshivtsev} have also been considered. The inverse fermion propagator in this case takes the form
\be
S_F^{-1}(p;\xi)=  A_e(p;\xi)\slsh{p} +A_o(p;\xi) \tau \slsh{p}-B_e(p;\xi)-B_o(p;\xi)\tau\;.
\ee
We explicitly label the propagator with the covariant gauge parameter $\xi$  
as we would be interested in its expression in different gauges.
The bare propagator corresponds to 
$A_e^{(0)}=1,\,A_o^{(0)}=0,\,B_e^{(0)}=m_e,\,B_o^{(0)}=m_o$.  In coordinate 
space, we have that
\be
S_F^{-1}(x;\xi)=  X_e(x;\xi)\slsh{x} +X_o(x;\xi) \tau \slsh{x}-Y_e(x;\xi)-Y_o(x;\xi)\tau\;.
\ee
Rather than working with parity eigenstates, we find it convenient to work in a chiral basis. 
For this purpose, we
introduce the chiral projectors $\chi_\pm = (1\pm \tau)/2 $ which have the properties 
$\chi_\pm^2=\chi_\pm\;, 
\chi_+\chi_-=0\;, \chi_++\chi_-=1$~\footnote{Further properties are shown in the Appendix.}. 
The right-handed $\psi_+$ and left-handed $\psi_-$ fermion fields are $\psi_\pm=\chi_\pm \psi$, 
in such a fashion that the chiral decomposition of the fermion propagator becomes
\bea
S_F(p;\xi)&=& -\frac{A_+ (p;\xi)\slsh{p}+B_+(p;\xi)}{A_+^2(p;\xi)p^2+
B_+^2(p;\xi)}\chi_+ 
- \frac{A_- (p;\xi)\slsh{p}+B_-(p;\xi)}{A_-^2(p;\xi)p^2+B_-^2(p;\xi)}\chi_- 
\nn\\
&\equiv& -[{\cal P}_+^V(p;\xi) \slsh{p}+{\cal P}_+^S(p;\xi)] \chi_+
-[{\cal P}_-^V(p;\xi) \slsh{p}+{\cal P}_-^S(p;\xi)] \chi_-\;,\label{fpp}
\eea
and analogously, in coordinate space
\bea
S_F(x;\xi)&=& -\frac{X_+ (p;\xi)\slsh{x}+Y_+(x;\xi)}{X_+^2(x;\xi)x^2+Y_+^2(x;\xi)}\chi_+
- \frac{X_- (x;\xi)\slsh{x}+Y_-(x;\xi)}{X_-^2(x;\xi)x^2+Y_-^2(x;\xi)}\chi_- \nn\\
&\equiv& -[{\cal X}_+^V(x;\xi) \slsh{x}+{\cal X}_+^S(x;\xi)] \chi_+ 
-[{\cal X}_-^V(x;\xi) \slsh{x}+{\cal X}_-^S(x;\xi)] \chi_-\;,\label{fpx}
\eea
where our notation is as follows: $K_\pm=K_e\pm K_o$ for $K=A,B,X,Y$, while ${\cal K}^V$ 
and ${\cal K}^S$, for ${\cal K} = {\cal P,X}$, stand for the vector and scalar  parts of 
the right- and left- projections of the fermion propagator, respectively.
Obviously, propagators~(\ref{fpp}) and~(\ref{fpx}) are related trough the Fourier 
transforms
\be
S_F(p;\xi)=\int d^3x e^{-ip\cdot x}S_F(x;\xi)\;, \qquad
S_F(x;\xi)= \int \frac{d^3p}{(2\pi)^3} e^{ip\cdot x} S_F(p;\xi)\;.\label{FT}
\ee
From these definitions, we are ready to study the LKFT for the fermion propagator, which  
we shall introduce in the next section, along with the strategy of its implementation 
for the study of gauge covariance of the fermion propagator.

\section{LKFT and the Non Perturbative Fermion Propagator }

The LKFT
relating the coordinate space fermion propagator in Landau gauge
to the one in an arbitrary covariant gauge in arbitrary spacetime 
dimensions $d$ reads~:
\bea
S_F(x;\xi) = S_F(x;0)e^{-i [\Delta_d(0)-\Delta_d(x)]} \;,
\eea
where
\bea
\Delta_d (x)=-\frac{i \xi e^2}{16 (\pi)^{d/2}}
(\mu x)^{4-d}\Gamma\left(\frac{d}{2}-2\right) \;,\label{deltad}
\eea
$\mu$ being a mass scale introduced for dimensional purposes. 
Explicitly in three dimensions,  the LKFT is given by
\bea
S_F(x;\xi)=e^{-ax}S_F(x;0)\label{LKFT}
\eea
where $a=\alpha\xi/2$ and $\alpha=e^2/(4\pi)$ as usual. With these definitions, 
we are ready to study the gauge covariance of the
 fermion propagator from its LKFT. The strategy is as follows~: (i) 
Start from the bare propagator in momentum space in Landau gauge and 
Fourier transform it to coordinate space. (ii) Apply the LKFT.
(iii) Fourier transform it back to momentum space.
We shall proceed to carry out this exercise below. Considering the bare 
propagator in  Landau gauge, 
we have $A_+^{(0)}(p;0)=A_-^{(0)}(p;0)\ =\ 1$ and
$B_\pm^{(0)}(p;0)= m_e\pm m_o\ \equiv \ m_\pm$. Therefore
\be
{\cal P}_\pm^{S\,(0)}(p;0) = \frac{m_\pm}{p^2+m_\pm^2}\;,  \qquad
{\cal P}_\pm^{V\,(0)}(p;0) = \frac{1}{p^2+m_\pm^2}\;.
\ee
Performing the Fourier transforms, we find
\be
{\cal X}_\pm^S(x;0)=\frac{m_\pm e^{- m_\pm x}}{4\pi x}\;,\qquad
{\cal X}_\pm^V(x;0)=\frac{i(1+m_\pm x)e^{-m_\pm x}}{4\pi x^3}\;.
\ee

The LKFT is straight forward to perform. It would merely shift the 
argument of the exponentials in the above expressions by the amount 
$-ax$. Then we are only left with the inverse Fourier transform, which 
leads to
\bea
{\cal P}_\pm^S(p;\xi)&=&\frac{m_\pm}{p^2+(a+m_\pm)^2}\nn\\
{\cal P}_\pm^V(p;\xi)&=&\frac{1}{p^2}\Bigg[1-\frac{m_\pm(a+m_\pm)}{p^2+(a+m_\pm)^2}
-a I(p,a+m_\pm) \Bigg]\;,
\label{LKFresults}
\eea
where we have defined
\be
I(p,m)=\frac{1}{p}\arctan{\left(\frac{p}{m} \right)}\;.\label{Ip}
\ee
The expressions (\ref{LKFresults}) yield the non perturbative form of the fermion propagator 
in an arbitrary covariant gauge. 
{\em An important advantage of the LKFT over ordinary perturbative 
calculation is 
that the weak coupling expansion of this transformation already fixes some of 
the coefficients in the all order perturbative expansion of the fermion 
propagator (see, for example,~\cite{LKFper,LKFperb,LKFperc,LKFper2}).}  
It is easy to show that
%
the coefficients of the terms of the form 
$\alpha^i\xi^i$ get already fixed in the all order perturbative expansion of 
the LKFT, starting from the bare propagator, a fact that holds true in arbitrary 
space-time dimensions, as pointed out in~\cite{LKFper2}. Even more, if we 
had started with a ${\cal O}(\alpha^n)$ propagator, all  the terms of the 
form $\alpha^{n+i}\xi^i$ would already get fixed, as well as those with higher 
powers of $\xi$ at a given order in $\alpha$ after the perturbative 
expansions of the results obtained on applying the corresponding LKFT.
Below we shall consider 
equation~(\ref{LKFresults}) in various limiting cases, for consistency checks.

\subsection{Massless case}

In the massless case, $m_e=m_o=0$, the non-perturbative fermion propagator reduces to
\be
{\cal P}^S_{\pm\,massless}(p;\xi)=0\;, \qquad
{\cal P}^V_{\pm\,massless}(p;\xi) = 
\frac{1}{p^2}\left[1- a I(p,a) \right]\;,
\ee
which imply $B_\pm(p;\xi)=0$ and hence 
$B_e(p;\xi)=B_o(p;\xi)=0$, i.e, fermions remain massless in all gauges. Furthermore, $A_+(p;\xi)=A_-(p;\xi)$, such that 
$A_o(p;\xi)=0$ and
\be
A_e(p;\xi)= \left[1-a I(p,a)\right]^{-1} \;,
\ee
confirming the covariant form for the massless propagator dictated by LKFT~\cite{LKFper,LKFperb}.

\subsection{Ordinary QED$_3$ case}

The ordinary, parity-conserving case was considered in~\cite{LKFperb}. It can be 
derived from our results setting $m_0=0$, which implies $m_\pm=m_e$. Hence 
we straightforwardly see that ${\cal P}_+^S(p;\xi)={\cal P}_-^S(p;\xi)$ and 
${\cal P}_+^V(p;\xi)={\cal P}_-^V(p;\xi)$, which in turn imply ${\cal P}^S_o(p;\xi)= 
{\cal P}^V_o(p;\xi)=0$ and thus we only have non vanishing contribution from the 
even-parity part of the fermion propagator:
\bea
{\cal P}_e^S(p;\xi)&=&\frac{m_e}{p^2+(a+m_e)^2}\nn\\
{\cal P}_e^V(p;\xi)&=&\frac{1}{p^2}\Bigg[1-\frac{m_e(a+m_e)}{p^2+(a+m_e)^2}
-a I(p,a+m_e) \Bigg]\;.
\label{LKFPC}
\eea
A comparison against the results of~\cite{LKFperb} shows complete agreement in this case.

\subsection{Weak Coupling Regime}

Next, we take the weak coupling limit of equation~(\ref{LKFresults}) 
performing an expansion of these expressions in powers of $\alpha$, 
recalling that $a=\alpha\xi/2$. At ${\cal O}(\alpha)$ we find
\bea
{\cal P}_{\pm\,weak}^S(p;\xi)&=&\frac{m_\pm}{p^2+m_\pm^2}- \frac{\alpha\xi 
m_\pm^2}{(p^2+m_\pm^2)^2}\nn\\
{\cal P}_{\pm\,weak}^V(p;\xi)&=&\frac{1}{p^2+m_\pm^2}+
\frac{\alpha\xi m_\pm(m_\pm^2-p^2)}{2p^2(p^2+m_\pm^2)^2}
-\frac{\alpha\xi}{2p^2} I(p,m_\pm)
\;.\label{lkfpert}
\eea
As we have pointed out earlier, the non perturbative expressions obtained from the LKFT of the 
fermion propagator matches onto perturbative results at the one-loop level up to 
a gauge independent term. In order to identify such a term, we need to calculate the
one-loop perturbative result of the propagator and compare against equation~(\ref{lkfpert}).
For this purpose it is better to work directly with the $A_\pm$ and $B_\pm$ 
functions, which at ${\cal O}(\alpha)$ are obtained from
\bea
A_\pm^{(1)}(p;\xi)&=&1
-\frac{2\pi\alpha}{p^2} \int \frac{d^3k}{(2\pi)^3} 
{\rm Tr} \Bigg[\slsh{p}\gamma_\mu S_\pm(k;\xi)\gamma_\nu 
\Delta_{\mu\nu}^{(0)}(q)\chi_\pm \Bigg]\;,\nn\\
B_\pm^{(1)}(p;\xi)&=&m_\pm
-2\pi\alpha \int \frac{d^3k}{(2\pi)^3} 
{\rm Tr}\Bigg[\gamma_\mu S_\pm(k;\xi)\gamma_\nu 
\Delta_{\mu\nu}^{(0)}(q)\chi_\pm \Bigg]\;,\label{pertraces}
\eea
where $q=k-p$ and 
$S_\pm(k;\xi)={\cal P}_\pm^{V\,(0)}(k;\xi) \slsh{k}+
{\cal P}_\pm^{S\,(0)}(k;\xi)$.
Using the explicit form of the bare photon propagator, i.e., 
$
\Delta_{\mu\nu}^{(0)}(q;\xi)=\left(q^2 \delta_{\mu\nu}+(\xi-1)
{q_\mu q_\nu} \right)/q^4$,
we find
\bea
A_\pm^{(1)}(p;\xi)&=&1
-\frac{\alpha\xi}{2\pi^2 p^2} \int d^3k 
\frac{(k^2+p^2)(k\cdot p)-2k^2p^2}{q^4(k^2+m_\pm^2)}\;,\nn\\
B_\pm^{(1)}(p;\xi)&=&m_\pm
+\frac{\alpha(2+\xi)m_\pm}{2\pi^2} \int d^3k 
\frac{1}{q^2(k^2+m_\pm^2)}\;.
\eea
These expressions are similar to the one-loop calculation carried out for the parity-even Lagrangian 
of QED$_3$ in~\cite{BaRa1}. The integration readily yields
\bea
A_\pm^{(1)}(p;\xi)&=& 1+\frac{a}{p^2}[m_\pm-(m_\pm^2-p^2)I(p,m_\pm)]\;,\nn\\
B_\pm^{(1)}(p;\xi)&=& m_\pm \left[1+\alpha(\xi+2)I(p,m_\pm) \right]\;.
\label{oneloop}
\eea
From the above expressions we can reconstruct ${\cal P}_\pm^{S\,(1)}$ 
and ${\cal P}_\pm^{V\,(1)}$, finding
\bea
{\cal P}_\pm^{S\,(1)}(p;\xi)&=& \frac{m_\pm}{p^2+m_\pm^2}-\frac{ \alpha\xi 
m_\pm^2}{(p^2+m_\pm^2)^2}
+\frac{2\alpha m_\pm(p^2-m_\pm^2)}{(p^2+m_\pm^2)^2}I(p,m_\pm) \;,\nn\\
{\cal P}_\pm^{V\,(1)}(p;\xi)&=&\frac{1}{p^2+m_\pm^2}+
\frac{\alpha\xi m_\pm (m_\pm^2-p^2)}{2p^2(p^2+m_\pm^2)^2}
-\frac{\alpha\xi}{2p^2}I(p,m_\pm) \nn\\&&
-\frac{4\alpha m_\pm^2}{p(p^2+m_\pm^2)^2}\;.
\label{pert}
\eea

Comparing these results against those obtained from LKFT, 
equation~(\ref{lkfpert}), we observe perfect agreement up to gauge 
independent terms, a difference allowed by the structure of LKFTs.
Note that in the Lagrangian~(\ref{lag}), only the term $m_o\bar\psi\tau\psi$ 
is parity odd. Such a term would radiatively induce a 
Chern-Simons term into the Lagrangian, modifying the form of the photon propagator. 
We study the extended Lagrangian in the following section.

\section{Maxwell-Chern-Simons QED$_3$}

 The fact that the parity-odd mass term in the fermion propagator radiatively induces a parity-odd 
contribution into the vacuum polarization can be seen from the tensor structure of the vacuum 
polarization $\Pi_{\mu\nu}(q)$ at the one-loop level 
\bea
\Pi_{\mu\nu}(q) &=& e^2 \int \frac{d^3k}{(2\pi)^3} {\rm Tr}\left[ \gamma_\mu S(k,\xi) \gamma_\nu S(k+q;\xi)\right]\nn\\
&=& \left( \delta_{\mu\nu}-\frac{q_\mu q_\nu}{q^2}\right) \Pi^e(q^2) + \epsilon_{\mu\nu\rho}q_\rho \Pi^o(q^2)\;.\label{Pi}
\eea
The second term corresponds to a Chern-Simons interaction of the form
\be
{\cal L}_{CS}=\frac{\theta}{4}\varepsilon_{\mu\nu\rho}A_\mu F_{\nu\rho}\;,\label{csl}
\ee
where $\theta=\Pi^o(q^2\to 0)$. This term is parity non invariant. Despite the fact that it is 
not manifestly gauge invariant, under a gauge transformation, ${\cal L}_{CS}$ changes by a total 
derivative (see for example~\cite{khare}),  rendering the corresponding action gauge invariant. 
The parameter $\theta$ induces a topological mass for the photons. Remarkably 
enough, Coleman and Hill~\cite{CH} demonstrated on very general grounds that this parameter 
receives no contribution from two- and higher-loops. Thus, it is desirable that in the presence 
of the parity violating mass term for the fermions in the Lagrangian, the Chern-Simons term 
should be considered as well.  The Maxwell-Chern-Simons QED$_3$ Lagrangian in this case takes the form
\be
{\cal L}= \bar\psi (i\slsh{\partial}+e\slsh{A}-m_e-\tau m_o)\psi
-\frac{1}{4}F_{\mu\nu}F_{\mu\nu}-
\frac{1}{2\xi}(\partial_\mu A_\mu)^2+\frac{\theta}{4}\varepsilon_{\mu\nu\rho}A_\mu F_{\nu\rho}.\label{mcslag}
\ee
This Lagrangian has been employed to describe the zero field quantum Hall effect for massive Dirac 
fermions~\cite{zero}. In that, the gauge invariant topological mass $\theta$ is found to be related to 
the Hall conductivity. Whichever modification this parameter should induce in the 
perturbative form of the fermion propagator, it certainly will {\em not modify the gauge dependence 
we found in the previous section. Thus equation~(\ref{LKFresults}) continues to be the same in the present case.}
In order to identify the role of the Chern-Simons term in the perturbative expansion of the fermion 
propagator, we first notice that the photon propagator associated to the Lagrangian (\ref{mcslag}) takes the form
\be
\Delta_{\mu\nu}^{(0)}(q;\xi)=\frac{1}{q^2+\theta^2}\left(\delta_{\mu\nu}-
\frac{q_\mu q_\nu}{q^2} \right)-\frac{\varepsilon_{\mu\nu\rho}q_\rho \theta}{q^2(q^2+\theta^2)}+\xi \frac{q_\mu q_\nu}{q^4}\;.
\ee
Inserting this propagator into (\ref{pertraces}) and taking traces, we have
\bea
A_\pm^{(1)}(p;\xi)&=&1
-\frac{\alpha\xi}{2\pi^2 p^2} \int d^3k 
\frac{(k^2+p^2)(k\cdot p)-2k^2p^2}{q^4(k^2+m_\pm^2)}\nn\\
&&+ \frac{\alpha}{\pi^2p^2} \int d^3k \frac{(k\cdot q)(p\cdot q)}{q^2 (q^2+\theta^2)(k^2+m_\pm^2)}\nn\\
&&\mp \frac{\alpha\theta m_\pm}{\pi^2 p^2}\int d^3k \frac{(p\cdot q)}{q^2 (q^2+\theta^2)(k^2+m_\pm^2)}
\;,\nn\\
B_\pm^{(1)}(p;\xi)&=&m_\pm
+\frac{\alpha\xi m_\pm}{2\pi^2} \int d^3k \frac{1}{q^2(k^2+m_\pm^2)}\nn\\&&
+\frac{\alpha m_\pm}{\pi^2}\int d^3k \frac{1}{(q^2+\theta^2)(k^2+m_\pm^2)} \nn\\
&&\mp \frac{\theta \alpha}{\pi^2}\int d^3k \frac{(k\cdot q)}{q^2(q^2+\theta^2)(k^2+m_\pm^2)}\;.
\eea
Using dimensional regularization, these integrals can be evaluated in a straightforward manner, yielding
\bea
&&\hspace*{-15mm}
A_\pm^{(1)}(p;\xi)\ =\ 1 + \frac{\alpha}{2p^2\theta^2}\Bigg\{ 
\left(\theta^2-p^2-m_\pm^2 \right)\left( \theta^2+p^2+m_\pm^2\pm 2m_\pm \theta \right)I(p,\theta+m_\pm)\nn\\
&&+\left[(p^2+m_\pm^2)(p^2+m_\pm^2 \pm 2m_\pm \theta)+\xi\theta^2(p^2-m_\pm^2) \right]I(p,m_\pm)\nn\\
&&+m_\pm\theta^2  (\xi+1\mp 2)-\theta\left(p^2+m_\pm^2+\theta^2\right)  \Bigg\}\;,\nn\\
&&\hspace*{-15mm}
B_\pm^{(1)}(p;\xi)\ =\ m_\pm+\frac{\alpha}{\theta}\Bigg\{ \left[2 m_\pm \theta \pm (p^2+m_\pm^2+\theta^2)\right]I(p,\theta+m_\pm)\nn\\
&&+\left[\xi m_\pm\theta \mp (p^2+m_\pm^2) \right]I(p,m_\pm)\pm\theta\Bigg\}\;.\label{1lCS}
\eea
Some particular limits of these expressions are considered below.

\subsection{Massless photons}

As $\theta\to 0$, we observe that
\bea
A_{\pm \ (\theta\to 0)}^{(1)}(p;\xi)&=& 1+\frac{a}{p^2}[m_\pm-(m_\pm^2-p^2)I(p,m_\pm)]\nn\\&&
-\frac{\alpha\theta}{3p^2}\left(2\pm  \frac{(3\mp2)m_\pm^2}{p^2+m_\pm^2} +3m_\pm I(p,m_\pm)\right)\;,\nn\\
B_{\pm \ (\theta\to 0)}^{(1)}(p;\xi)&=& m_\pm \left[1+\alpha(\xi+2)I(p,m_\pm) \right]\nn\\&&
+\alpha\theta \left(-\frac{(2\mp1)m_\pm}{p^2+m_\pm^2}\pm I(p,m_\pm)\right).\label{Cs0}
\eea
A comparison against (\ref{oneloop}) reveals that we recover the ``pure'' QED$_3$ limit when 
photons are massless, i.e, $\theta=0$.

\subsection{Massless fermions}

In the absence of the Maxwell-Chern-Simons term, equation (\ref{oneloop}) reveals that if we start
from massless fermions, i.e., $m_\pm=0$, radiative corrections, being proportional to the bare mass, 
do not alter their masslessness. However, for $m_\pm=0$ in the present case, we see from~(\ref{1lCS})  that
\be
B_{\pm (m_\pm=0)}^{(1)} (p;\xi)= \pm \frac{\alpha}{\theta}\Bigg[-\frac{\pi p}{2}+\theta + (p^2+\theta^2)I(p,\theta) \Bigg]\;,
\ee
which readily implies $B_e(p;\xi)=0$, but $B_o(p;\xi)\propto \alpha/\theta$. This implies that even 
starting with massless fermions, the Maxwell-Chern-Simons term radiatively induces a parity violating 
mass for them. In fact, we can see that in the Landau gauge, as $\theta\to 0$, the  
induced mass function is,
\be
m_o^{induced}(p;0)=\lim_{\theta\to 0}\frac{B_{\pm (m_\pm=0)}^{(1)} (p;0)}{A_{\pm (m_\pm=0)}^{(1)} (p;0)} = \frac{\alpha\theta\pi}{2p}\;,
\ee
and would be zero if we turn off either the interactions, i.e., $\alpha=0$, or the Maxwell-Chern-Simons mass, $\theta=0$.
Such a statement, complementary to the Coleman-Hill theorem~\cite{CH}, was first noticed in~\cite{waites}, 
and stresses the need to include in the bare Lagrangian both the Maxwell-Chern-Simons term and the Haldane mass 
term simultaneously, or none at all. 

\subsection{Ordinary QED$_3$ case}

The ordinary QED$_3$ case is recovered by setting $\theta=m_o=0$ in (\ref{1lCS}). This 
can be achieved in two steps: First, from (\ref{Cs0}) we recover the pure QED$_3$ results 
(\ref{oneloop}) by setting $\theta=0$. We then arrive at (\ref{LKFPC}) 
by setting $m_o=0$ in (\ref{oneloop}), as we have previously pointed out.

\subsection{Gauge dependent terms}

In order to perform a comparison against the perturbative expansion of the LKFT 
results,~equation~(\ref{lkfpert}), it is convenient to return to the scalar 
and vector parts of the propagator. We find that
\bea
{\cal P}_{\pm}^S(p;\xi)&=&\frac{m_\pm}{p^2+m_\pm^2}- \frac{\alpha\xi 
m_\pm^2}{(p^2+m_\pm^2)^2}\nn\\
&+&\frac{\alpha}{\theta^2(p^2+m_\pm^2)^2}\Bigg\{ \theta \left[m_\pm (p^2+m_\pm^2+\theta^2)\pm \theta(p^2+(1\mp 1)m_\pm^2) \right]\nn\\
&+&  \left(p^2+(\theta \pm m_\pm)^2 \right)
\Bigg[p^2(\theta\pm m_\pm)+m_\pm^2(m_\pm \mp \theta) 
-\theta^2m_\pm \Bigg] \nn\\
&\times& I(p,\theta+m_\pm)-(p^2+m_\pm^2)^2 (\theta\pm m_\pm)I(p,m_\pm)\Bigg\}\;,\nn\\
{\cal P}_{\pm}^V(p;\xi)&=&\frac{1}{p^2+m_\pm^2}+
\frac{\alpha\xi m_\pm(m_\pm^2-p^2)}{2p^2(p^2+m_\pm^2)^2}
-\frac{\alpha\xi}{2p^2} I(p,m_\pm)\nn\\
&+&\frac{\alpha}{2p^2\theta^2(p^2+m_\pm^2)^2} \Bigg\{ 
\theta \left[(p^2-m_\pm^2)\left(p^2+m_\pm^2+\theta^2\right)\right.\nn\\
&\mp& \left.\theta m_\pm \left((2\pm1)p^2 + (2\mp 1) m_\pm^2 \right)\right]\nn\\
&+&\left(p^2+(\theta\pm m_\pm)^2\right) \left[(p^2+m_\pm^2-\theta^2)(p^2-m_\pm^2)\mp 4 p^2 \theta m_\pm  \right]\nn\\ &\times &I(p,\theta+m_\pm)- (p^2+m_\pm^2)^2(p^2-m_\pm^2\mp 2\theta m_\pm) I(p,m_\pm)\Bigg\}\;.\label{CSpert}
\eea 
The gauge dependent terms exactly match onto the LKFT results expanded in the weak coupling limit, 
as expected. Furthermore, the gauge independent terms, as compared to those in equation~(\ref{pert}) 
exhibit a more intricate dependence on the topological parameter $\theta$. These cannot be derived 
from the LKFT of the tree-level fermion propagator alone.

\subsection{Numerical results}

In perturbation theory, higher order terms in the expansion parameter 
$\alpha$ are smaller than the lower order terms. Naturally, one wonders 
about how far would be the one-loop result  as compared to 
the non perturbative one obtained from the LKFT in quantitative terms. In 
Figure~1, we have drawn the scalar and vector projections of the fermion 
propagator propagator in various gauges arising from: non perturbative LKF analysis,
equation~(\ref{LKFresults}) and the one-loop perturbative treatment, equation~(\ref{CSpert}). 
The additional gauge parameter independent terms in the one-loop 
results, which are absent in the weak coupling expansion of the LKFT expressions,
seem to play a noticeable role in the infra red.
With increasing momentum, their contribution diminishes as both the expressions in 
equation~(\ref{LKFresults}) and equation~(\ref{CSpert}) start merging into each other, a statement 
that seems to hold true in arbitrary covariant gauges.

\begin{figure}[t]
\begin{center}
\includegraphics[width=0.3\textwidth,angle=-90]{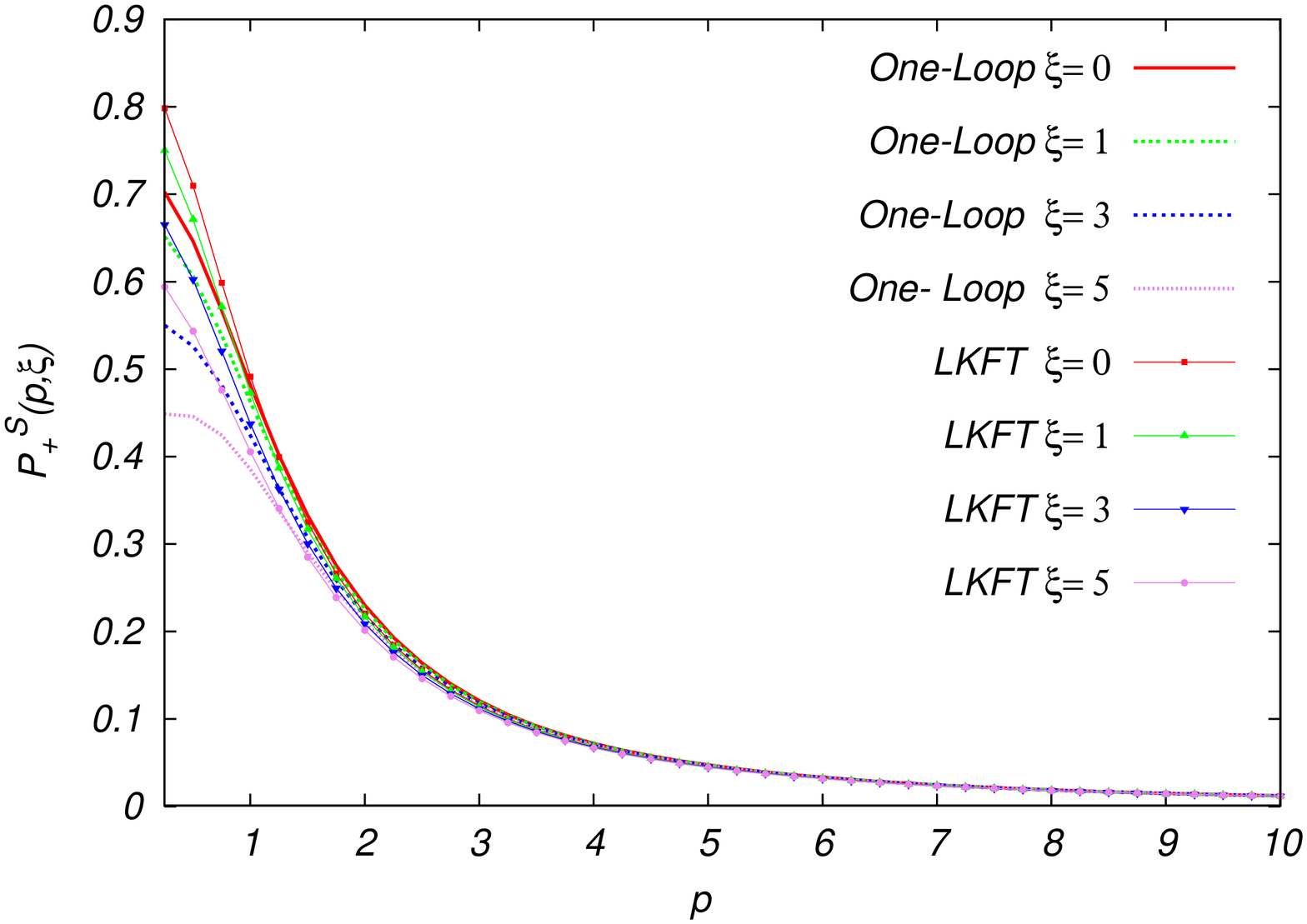} 
\includegraphics[width=0.3\textwidth,angle=-90]{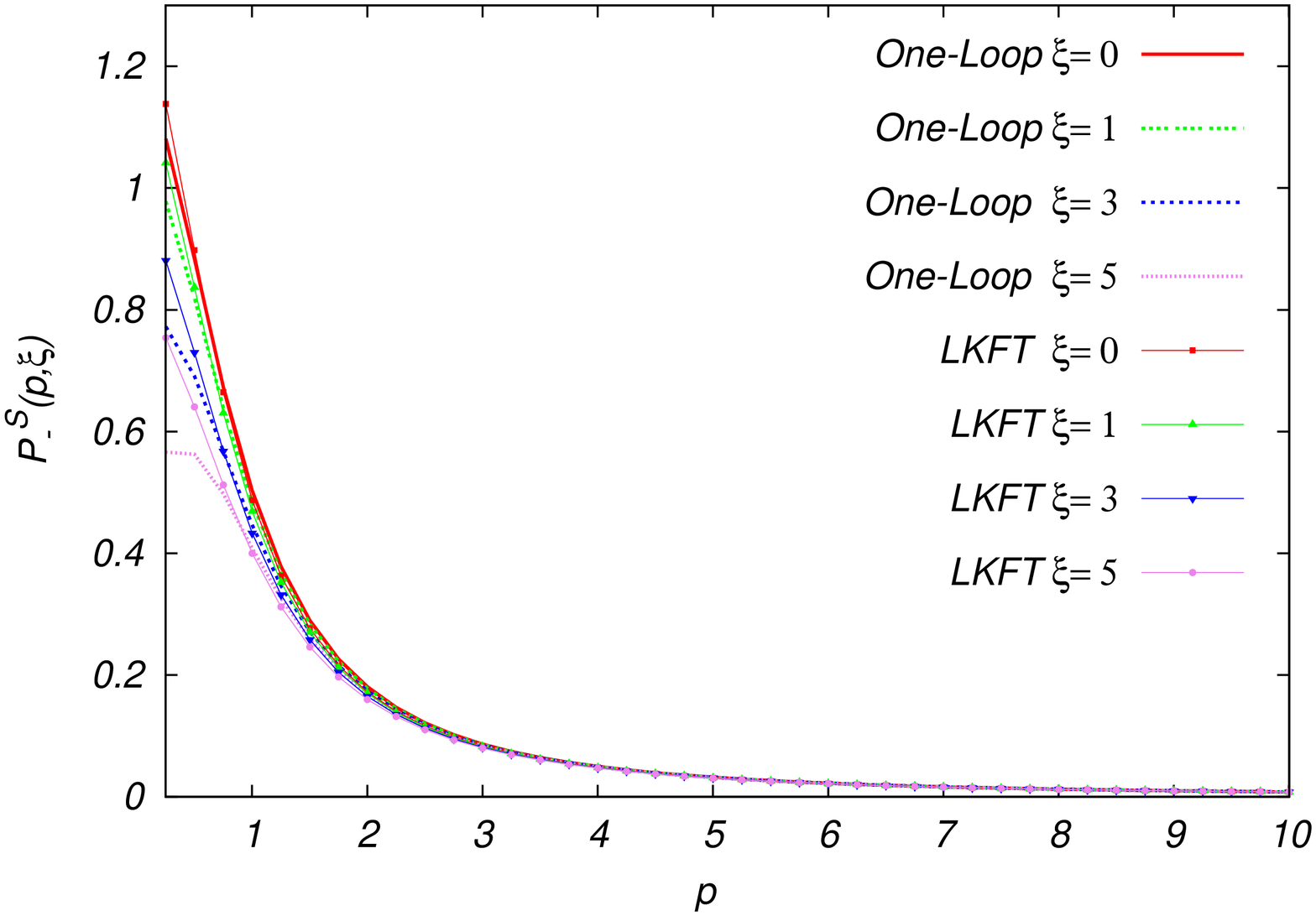} 

\includegraphics[width=0.3\textwidth,angle=-90]{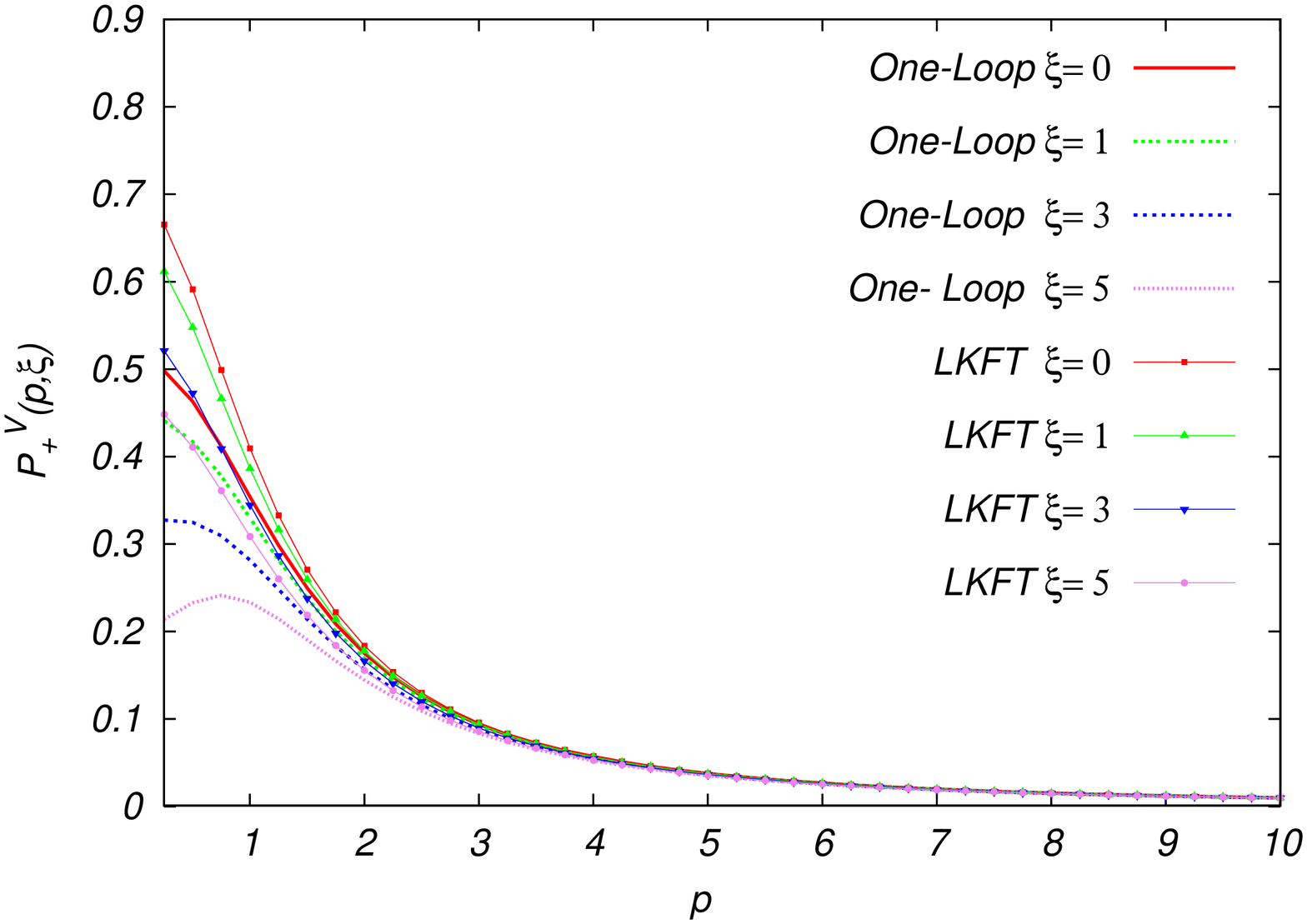} 
\includegraphics[width=0.3\textwidth,angle=-90]{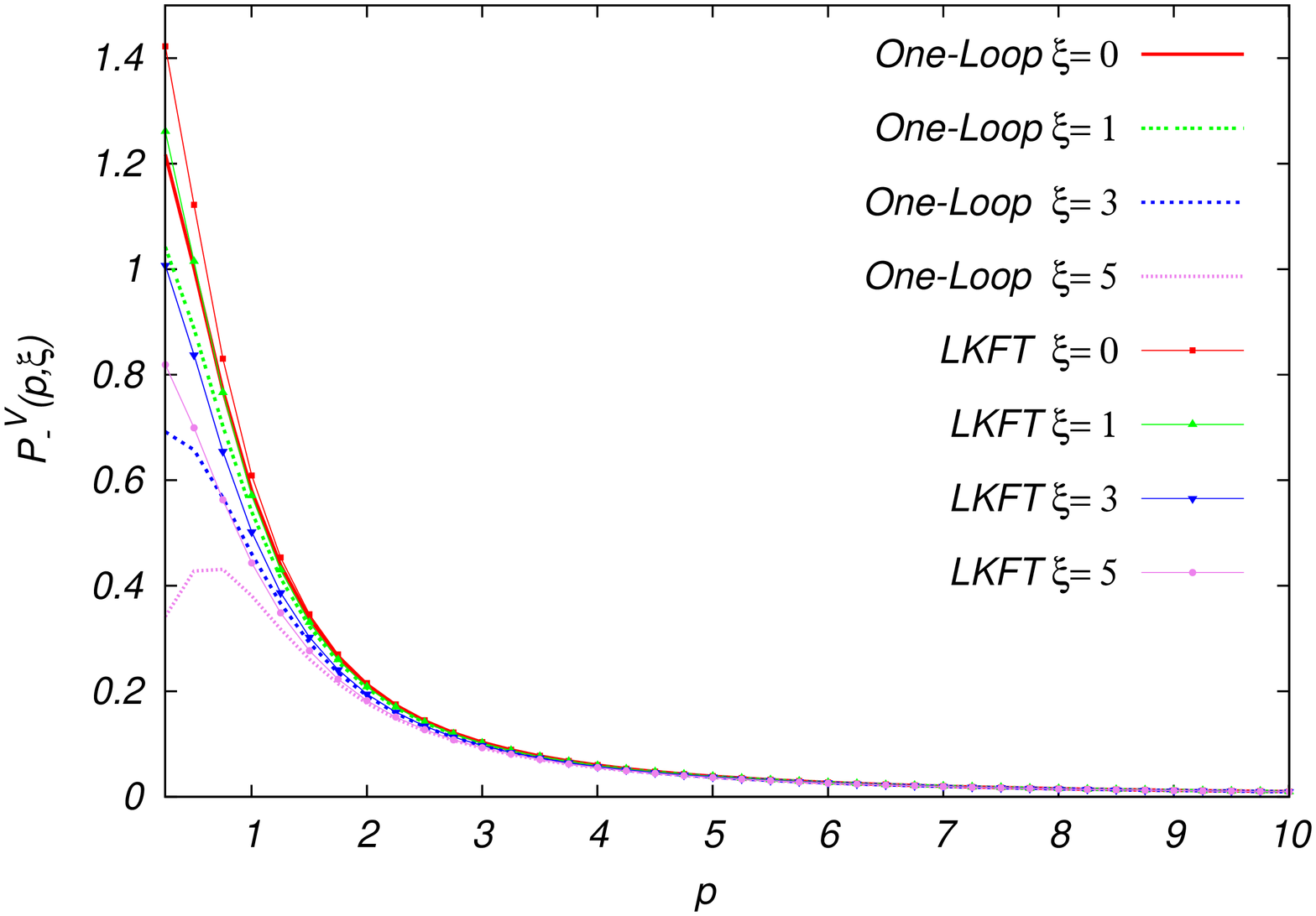} 
\end{center}
\caption{Scalar and vector projections of the fermion propagator in various 
gauges. The scale is set by the value of $e^2=1$ and we have chosen $m_e=1$, $m_o=0.2$ and $\theta=0.4$.}
\label{fig1}
\end{figure}

\section{Conclusions}

We have derived a non-perturbative expression for the fermion propagator in Maxwell-Chern-Simon QED$_3$ 
through its LKF transformation, starting from its tree level expression. 
Equation~(\ref{LKFresults}) displays one of the main results of this paper. The LKFT of the fermion 
propagator is written entirely in terms of basic functions of momentum, parity-even and parity-odd bare masses.  
Although our input is merely the bare propagator, its  LKFT, being non-perturbative in nature, 
contains useful information of higher orders in perturbation theory. All the coefficients of
the $(\alpha \xi)^i$ at every order are correctly reproduced without ever having to perform
loop calculations. In the weak coupling regime, 
LKFT results match onto the one-loop perturbative results derived from the Lagrangian~(\ref{lag}) 
up to gauge independent terms, a difference allowed by the structure of the LKFT. 
This difference arises due to our approximate input, and can be systematically removed at the 
cost of employing a more complex input which would need to be calculated by the brute
force of perturbation theory.

\ack
The authors  wish to thank the SNI, COEYCyT, CIC, and CONACyT grants under projects 4.10, 4.22 and 46614-I.

\appendix
\section*{Appendix }
\setcounter{section}{1}

The following trace identities are fulfilled by the $\chi_\pm$ projectors:
\bea
{\rm Tr}[\chi_\pm]&=&2\nn\\
{\rm Tr}[\gamma^\mu \chi_\pm]&=& 0\nn\\
{\rm Tr}[\gamma^\mu \gamma^\nu\chi_\pm]&=&-2 \delta^{\mu\nu}\nn\\
{\rm Tr}[\gamma^\mu\gamma^\nu\gamma^\alpha\chi_\pm]&=&\mp 2\epsilon^{\mu\nu\alpha}\nn\\
{\rm Tr}[\gamma^\mu\gamma^\nu\gamma^\alpha\gamma^\beta\chi_\pm]&=& 
2(\delta^{\mu\nu}\delta^{\alpha\beta}-\delta^{\mu\alpha}\delta^{\nu\beta}+\delta^{\mu\beta}\delta^{\nu\alpha}).
\eea

\end{document}